\documentclass[aps,prb,twocolumn,showpacs,10pt]{revtex4-1}
\usepackage{graphicx,amsmath,amssymb,amsfonts,sidecap,color}

\newcommand{\ket}[1]{\ensuremath{\left|#1\right\rangle}}

\newcommand{\ve}{\mathbf}

\makeatletter
\newcommand*{\rom}[1]{\expandafter\@slowromancap\romannumeral #1@}
\makeatother

\begin{document}

\title{Spintronics in ${\rm MoS}_2$ monolayer quantum wires}

\author{Jelena Klinovaja}
\affiliation{Department of Physics, University of Basel,
             Klingelbergstrasse 82, CH-4056 Basel, Switzerland}
\author{Daniel Loss}
\affiliation{Department of Physics, University of Basel,
             Klingelbergstrasse 82, CH-4056 Basel, Switzerland}

\date{\today}



\begin{abstract}
We study analytically and numerically spin effects in ${\rm MoS}_2$ monolayer  armchair quantum wires and quantum dots.  
The interplay between intrinsic  and Rashba spin orbit interactions induced by an electric field leads to helical modes, 
giving rise to spin filtering
in time-reversal invariant systems. The Rashba spin orbit interaction can also be generated by  spatially varying magnetic fields. In this case, the system can be in a helical regime with nearly perfect spin polarization.  
If such a quantum wire is brought into proximity to an $s$-wave superconductor, the system can be tuned into a topological phase, resulting in midgap Majorana fermions
localized at the wire ends.
\end{abstract}

\maketitle

{\it Introduction.} Atomic monolayers such as graphene sheets \cite{Novoselov_2009} have attracted much attention over the years. However, the small spin orbit interaction (SOI) in graphene makes spin effects negligibly small. \cite{kane_mele,cnt_ext_kuemmeth,klinovaja_cnt,cnt_helical_2011,kane_mele,izumida,fabian} In contrast,  transition-metal dichalcogenide semiconductors, \cite{frindt_mos,Morrison_mos, mos_nanotubes, mos_ribbons,mos_Fuhrer,mos_gap,mos_ribbons_defects,mos_transistor, mos_etching,mos_steele,mos_sc,MOS_review} in particular ${\rm MoS}_2$,
possess giant values of SOI. \cite{MOS_review} 
Combined with a direct band gap this SOI makes these materials attractive for optical effects.
\cite{monolayer_optics_exp_2010, Zeng_optics_exp_2012,optics_Nature,Yao_2012,Niu_valley_Hall_2007,Niu_optics_rules_2008}
However, previous work emphasized valleytronics in ${\rm MoS}_2$, \cite{Niu_valley_Hall_2007,Niu_optics_rules_2008} while the spin degrees of freedom have received much less attention, despite the fact that these materials 
can be expected to display interesting spintronics effects, such as helical states in quantum wires,  Majorana fermions, spin qubits in quantum dots, electrical control of spin, {\it etc.} 
This gap of understanding has motivated the present work where we will propose and analyze spin effects specifically for quantum confined structures in ${\rm MoS}_2$ monolayers.

One of our main findings is that the  intrinsic SOI needs to be complemented by Rashba-like SOI to obtain interesting spin effects in the conduction band.
In particular, we will focus on suitably defined quantum wires of armchair type and show that they allow for helical modes, with and without time-reversal symmetry. The Rashba-like interaction can be generated
by breaking structure inversion symmetry with gates or adatoms or, alternatively, by nanomagnets with alternating magnetization direction.
Helical modes serve as basis for spin filters~\cite{streda} but also as platform for exotic quantum states such as Majorana fermions~\cite{alicea_review_2012} or fractionally charged fermions. \cite{Two_field_Klinovaja} We finally discuss quantum dots with well-defined Kramers doublets that can serve as spin qubits.~\cite{kloeffel_prospects_2013}

\begin{figure}[!bp]
    \centering
    \includegraphics[width=\columnwidth]{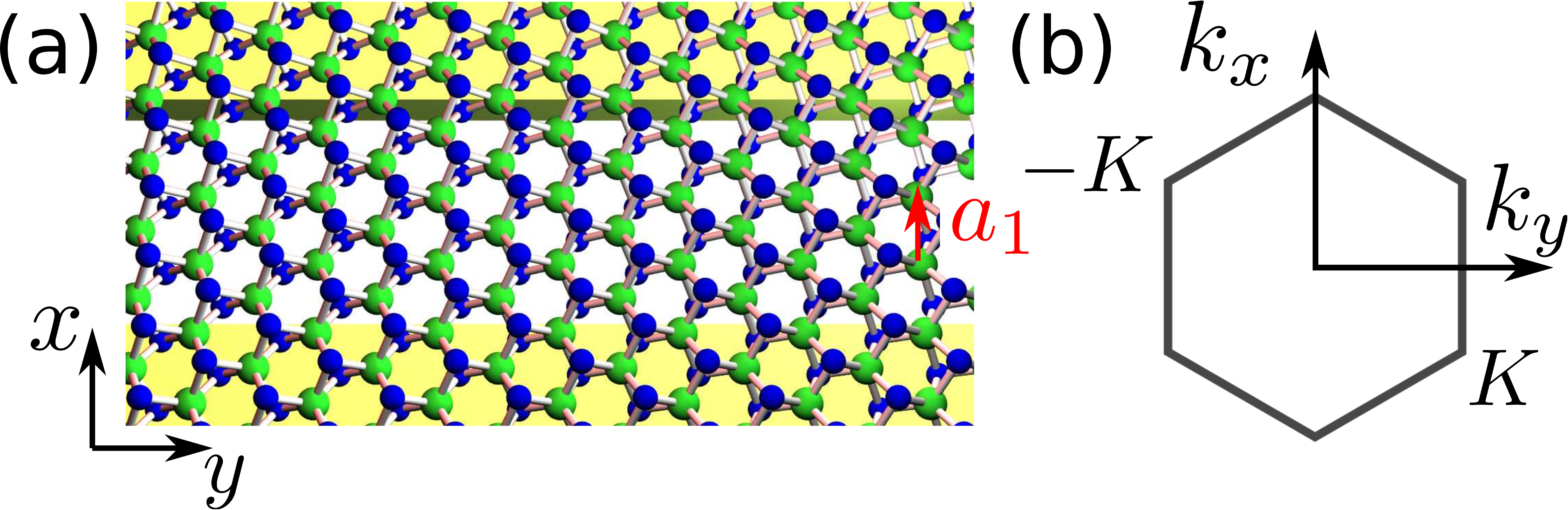}
\caption{ (a) The ${\rm MoS}_2$ monolayer lattice consists of  ${\rm Mo}$ (large green dots) each connected to six ${\rm S}$ (small blue dots).
The armchair quantum wire in the  monolayer can be formed by metallic gates (yellow area) that fix the propagation direction (defined as the $y$ axis) to be perpendicular to one of the  lattice translation vectors, say $\bf a_1$ (red arrow). 
(b) Brillouin zone where the valleys $K$ and $-K$ lie on the $k_x$ axis which is perpendicular to the direction of propagation given by the $k_y$ axis.
Note that the boundaries of a ${\rm MoS}_2$ flake are not important for this setup.}
\label{fig:mos}
\end{figure}

\begin{figure*}[!tp]
    \centering
    \includegraphics[width=1.75\columnwidth]{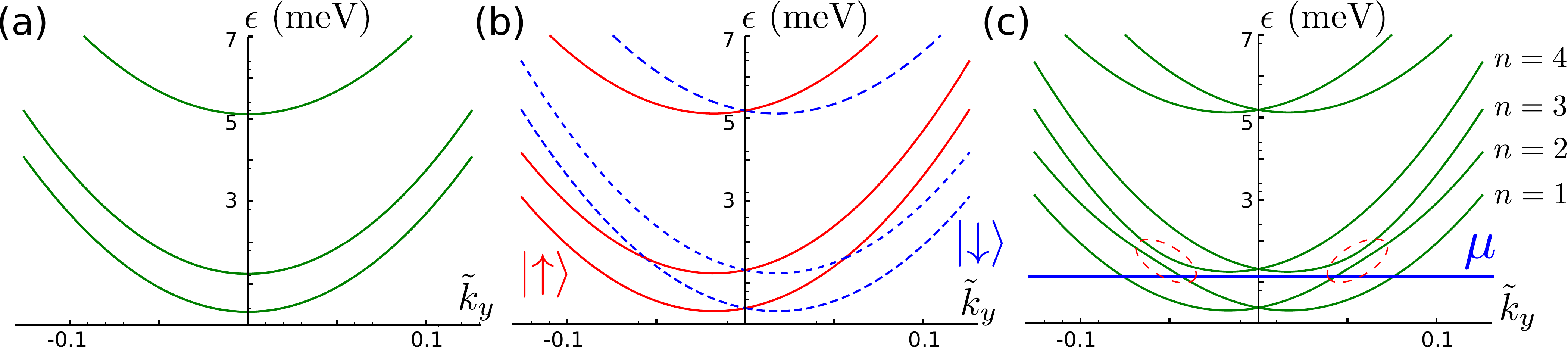}
\caption{(a) The energy spectrum of $H_0$ [$\epsilon(\tilde k_y)\equiv E(\sqrt{3} k_y a)-\Delta$] for a quantum wire of width $W=50a$ as obtained by numerical diagonalization. Parameters are chosen as $t=1.27\ {\rm eV}$, $\Delta=0.83\ {\rm eV}$, and $a=0.32\ {\rm nm}$. All levels are degenerate  only  in spin. 
This spectrum is in good agreement with our analytical predictions, see Eq. (\ref{spectrum_0}). (b) The Rashba SOI term $H_{Rx}$,  $\alpha_R=10 \ {\rm meV}$, lifts the spin-degeneracy and results in the spin-dependent shift of the wavevector $\tilde k_y$. (c) The remaining SOI terms: intrinsic SOI $H_{so}$, $\alpha=38 \ {\rm meV}$, and  Rashba SOI $H_{Ry}$, $\alpha_R=10 \ {\rm meV}$, lead to the anticrossings in the spectrum (red dashed circles).
}
\label{fig:spectrum_0}
\end{figure*}

{\it Bandstructure.} A molybdenum disulphide (${\rm MoS}_2$) monolayer consists of two layers of ${\rm S}$ atoms stacked over each other forming an effective trigonal lattice and of $\rm Mo$ atoms located in the center of the sulphur lattice, see Fig.~\ref{fig:mos}a. The Brillouin zone consists of a hexagon with two nonequivalent corners (valleys) at $\ve K$ $(\tau_z =1)$ and $ -\ve K$ $(\tau_z =-1)$ (see Fig.~\ref{fig:mos}b) that determine the low energy spectrum of the monolayer. \cite{Eriksson_2009_cones, Yao_2012, dft_mos_2012,Ataca_chemistry,kuc_2011}  This part of the spectrum is dominated by three $d$ orbitals of ${\rm Mo}$: the conduction band by $\ket{\psi_c}=\ket{d_{z^2}}$ and the valence band by $\ket{\psi_v^{\tau_z}}=(\ket{d_{x^2 - y^2}} + i \tau_z \ket{d_{xy}})/\sqrt{2}$. 
The effective Hamiltonian is given by
\begin{align}
H_0 = \hbar \upsilon_F (k_x \tau_z \sigma_1 + k_y \sigma_2) + \Delta \sigma_3,
\end{align}
where the Pauli matrices $\sigma_i$ act on the $d$ orbital space. The momenta $k_x$ and $ k_y$ are calculated from the corresponding valley characterized by $\tau_z$. Here, 
$\upsilon_F \approx 0.53 \times 10^6\ {\rm m/s}$  is the Fermi velocity and the mass term, $\Delta \approx 830\ {\rm meV}$, arising from broken inversion symmetry, has been extracted from DFT calculations.
\cite{Yao_2012} The spectrum of $H_0$ is given by
$E_{c,v}=\pm\sqrt{(\hbar \upsilon_F )^2(k_x^2+ k_y^2)+ \Delta^2}$.
The large gap $2\Delta$ between the valence and conduction bands makes the monolayer attractive for optical effects.
\cite{monolayer_optics_exp_2010, Niu_optics_rules_2008, Zeng_optics_exp_2012, optics_Nature,Yao_2012} The wavefunctions at fixed energy $E$ and momentum $k_y$ are written in the basis $(\psi_c, \psi_v^{\tau_z})$ as
\begin{equation}
\psi_{\tau_z, p}(x) = e^{i(\tau_z K+pk_x)x+ i k_yy} \begin{pmatrix}
                             b_{\tau_z, p}\\1
                            \end{pmatrix}
                            _{\hspace{-2.5pt}  {\tau_z}}, 
                            \label{fundamental}
                            \end{equation}
where $b_{\tau_z, p} = \hbar \upsilon_F (\tau_z p k_x - i k_y)/({E-\Delta})$, $k_x>0$, and the index $p$ labels right- and left- movers in $x$ direction.

We focus  now on the quasi-one-dimensional limit where the system forms a quantum wire.
Similarly to carbon-based materials, \cite{nanoribbon_KL,bilayer_MF_2012}
this quantum regime can be achieved in two ways: either by growing a nanoribbon of $\rm MoS_2$ with particular boundaries or by electrostatically confining the electrons  into a quantum wire by placing metallic gates on a $\rm MoS_2$ monolayer flake (with unspecified boundaries), see Fig. \ref{fig:mos}. In both cases we consider the {\em armchair} regime where the direction of propagation is perpendicular to a lattice translation vector. The details of the boundaries are not essential  provided that they do not suppress the transport. In addition, we assume  for both cases hard-wall type boundary conditions.
A most characteristic feature of such armchair quantum wires is that  the two valleys [$\pm \ve K = (\pm 4\pi/3a,0)$], projected onto the propagation direction along the $y$ axis,  coincide at $k=0$. 
The valleys easily hybridize (lifting their degeneracy), for instance, by impurities, irregular boundaries or, in particular, by our hard-wall boundaries.  \cite{brey_2006,nanoribbon_KL}

We determine now the spectrum for a quantum wire of width $W=Na$, where $a$ is the lattice constant and $N$ the number of unit cells in the $x$ direction.
To find the quantization conditions on $k_x$, we virtually extend our quantum wire by two sides, $W'=(N+2)a$, and impose the  boundary conditions  at these virtual sites on the total wavefunction $\psi(x)=\sum_{\tau_z,p}a_{\tau_z,p}\psi_{\tau_z, p}(x)\equiv \sum_{j}\psi_{j}(x)$, where $\psi_{j}(x)$ is the probability amplitude to find the electron in one of the orthogonal  states $j=d_{
z^2}, d_{x^2 - y^2}, d_{xy}$. This 
gives us $six$ conditions (three at each edge) for four fundamental solutions [see Eq.~(\ref{fundamental})]. To solve this overconstrained boundary value problem we use mixed boundary conditions.
On the orbitals $d_{z^2}$ and $d_{x^2 - y^2}$ we impose Dirichlet boundary conditions,  $\psi_{d_{z^2},d_{x^2 - y^2}} (x=0,W')=0$,
 while on the orbital $d_{xy}$ we impose von Neumann boundary conditions, $\partial_x\psi_{d_{xy}}(x=0,W')=0$.

These boundary conditions can be fulfilled only for those values of $k_x = |\kappa_m|$ that satisfy  $(K + \kappa_m)W' = \pi m$, where $m$ is an integer. If $W=(3M+1)a$, where $M$ is a positive integer, this leads to $\kappa_m= \pi m/W'$. We note that in this case all energy levels except the lowest one in the conduction band and the highest one in the valence band are two-fold degenerate.
If $W=(3M+2)a$  [$W=3Ma$], this leads to $\kappa_m= (\pi m + 2\pi/3) /W'$ [$\kappa_m= (\pi m - 2\pi/3) /W'$]. In this case, the energy levels are non-degenerate. The conduction band spectrum for small momenta is quadratic,
\begin{equation}
E_{c,w} = \sqrt{(\hbar \upsilon_F \kappa_m)^2+(\hbar \upsilon_F k_y)^2+ \Delta^2} \approx \Delta_m+\frac{(\hbar \upsilon_F k_y)^2}{2\Delta_m},
\label{spectrum_0}
\end{equation}
where we define the minimum energy for the $m$th subband  as $\Delta_m= \sqrt{(\hbar \upsilon_F \kappa_m)^2 +  \Delta^2}$, see Fig. \ref{fig:spectrum_0}a. We note that the slope of the spectrum branches at small momenta is decreasing with increasing $m$, resulting in crossings of different subbands, see Fig. \ref{fig:spectrum_0}c. The corresponding wavefunction is given by
\begin{align}
\psi_m(x) = \psi_{1, p}(x) - \psi_{-1, -p}(x),
\label{wavefunction_0}
\end{align}
where 
$ p = {\rm sgn} \kappa_m$. Again, we note that for $W=(3M+1)a$ the subbands $m$ and $-m$ have the same energy.

To confirm our analytical results numerically, we develop a tight-binding model for a honeycomb lattice composed of two kinds of atoms, $A$ (representing $d_{z^2}$ orbitals) and $B$ (representing $d_{x^2 - y^2}+i\tau_z d_{xy}$ orbitals). The effective Hamiltonian $\bar H_0$ consists of an on-site energy term and a term describing hopping between nearest neighbours, respectively,
\begin{equation}
\bar H_0 = \sum _i \varepsilon_i c_{i\mu}^\dagger c_{i\mu} + \sum_{<ij>} t_{ij} c_{i\mu}^\dagger c_{j\mu},
\end{equation}
where $c_{i\mu}^\dagger$  creates an electron with spin $\mu$ at site $i$.
The on-site energy $\varepsilon_i$ is equal to $\Delta$ ($-\Delta$) on $A$ ($B$) sublattice. The hopping matrix element $t_{ij}$ is assumed to be uniform, $t_{ij}\equiv t =2 \hbar \upsilon_F /\sqrt{3} a $. The Hamiltonians  $H_0$ and $\bar H_0$ are equivalent for momenta close to $\pm\ve K$ and result in the same low energy spectrum, see Fig. \ref{fig:spectrum_0}a.

{\it Intrinsic spin orbit interaction.} The intrinsic spin orbit interaction in $\rm MoS_2$ monolayer is much larger than in other monolayers, for example, in graphene,  arising from $d$ orbitals of the heavier atom. Symmetry arguments confirmed by DFT calculations lead to the intrinsic SOI Hamiltonian of the form \cite{Yao_2012}
\begin{equation}
H_{so} = \alpha \tau_z s_z (1-\sigma_3),
\label{h_so}
\end{equation}
where Pauli matrices $s_i$ act on spin space, and  $\alpha = 38\ {\rm meV}$ is the SOI strength.

{\it Rashba spin orbit interaction.} Breaking of structure inversion symmetry by an electric field $\ve E$ along the $z$ axis perpendicular to the monolayer leads to  a Rashba term of the form \cite{kane_mele} $H_R = H_{Rx} +H_{Ry}$,
\begin{equation}
H_{Rx} = -\alpha_R s_x \sigma_2,\ \ H_{Ry} = \alpha_R  \tau_z s_y \sigma_1,
\label{Rashba}
\end{equation}
where the Rashba SOI strength, in general, is proportional to the electric field strength. Such an electric field could be produced by gates \cite{kane_mele} or by doping with adatoms. \cite{Franz_2012, Rashba_2012} For both cases, $\alpha_R$ is best determined by {\it ab initio} calculations or experimentally.

An alternative way to generate Rashba SOI is to apply a spatially varying magnetic field, \cite{Braunecker_Jap_Klin_2009} produced, for instance, by nanomagnets. \cite{exp_field, Flensberg_Rot_Field} For example, a magnetic field ${\bf B}_n$ rotating in the  plane of a quantum wire produces the  Zeeman term
\begin{equation}
H_Z^\parallel = \Delta_Z [s_x \cos (k_n y) + s_y \sin (k_n y)],
\end{equation}
where $\Delta_Z = g\mu_B B_n/2$, and  the period of the rotating field is $\lambda_n = 2\pi/k_n$. The unitary spin-dependent transformation $U_n= \exp(-i k_n y s_z/2)$ allows us to gauge away the coordinate dependent term $H_Z^\parallel$ in the Hamiltonian $H=H_0+H_{so}+H_Z^\parallel$. This results in $H_n= U_n^\dagger H U_n$,
\begin{align}
H_n=H_0+H_{so}-\alpha_{Rn} s_z \sigma_2+\Delta_Z s_x,
\end{align}
where $\alpha_{Rn} = \hbar \upsilon_F k_n/2$, so the strength of the induced Rashba SOI depends only on the rotation period $\lambda_n$ but not on the magnetic field strength $B_n$. We note that  the induced Rashba SOI described by $H_{Rn}=-\alpha_{Rn} s_z \sigma_2$ reaches  $\alpha_{Rn}\approx 11\ {\rm me V}$ for the nanomagnets placed with a period $\lambda_n=100\ {\rm nm}$. 
Here we note that for a quantum wire created by gates we can estimate that the misalignment angle should be less than $a/\lambda_n$ (i.e. $ \lesssim 1 ^\circ$). \cite{bilayer_MF_2012}
If one works with a nanoribbon, then the propagation in the armchair or zigzag direction should be favoured by the growth process. \cite{nanoribbon_production, nanoribbon_production_CNT}
We note that also a Zeeman term $H_Z = \Delta_Z s_x$,  which breaks the time-reversal invariance of the system, inevitably arises.

To account for the Rashba SOI of Eq. (\ref{Rashba}) in the tight-binding model,  we allow for spin-flip hoppings,  \cite{kane_mele,nanoribbon_KL}
\begin{equation}
\bar{H}_{R}=\frac{3i \alpha_R}{4}\sum_{<ij>, \mu, \mu'}  c_{i\mu}^\dagger ({\boldsymbol e}_{ij} \times {\boldsymbol e}_{z} )\cdot {\ve s}_{\mu \mu'} c_{j\mu'},
\label{tb_soi}
\end{equation}
where  ${\ve s} = (s_x, s_y, s_z)$, 
and where the unit vectors ${\boldsymbol e}_{z}$ points along $z$ and ${\boldsymbol e}_{ij}$ along the bond connecting sites $i$ and $j$.
The intrinsics SOI, see Eq. (\ref{h_so}), can be modeled by
\begin{equation}
\bar{H}_{so}= \frac{2i\alpha }{3\sqrt{3}}\sum_{\ll ij\gg, \mu, \mu'}  \nu_{ij} c_{i\mu}^\dagger s_{z, \mu \mu'} c_{j\mu'},
\end{equation}
where the sum runs over the next-nearest neighbour sites belonging to the $B$ sublattice. The spin dependent amplitude $\nu_{ij} = -\nu_{ji}=\pm1$ depends on whether the electron takes a right or left turn by hopping from $i$ to $j$.  \cite{kane_mele} These two terms $\bar{H}_{so}$ and   $\bar{H}_{R}$ are constructed in such a way that they  are equivalent to $H_{so}$  and ${H}_{R}$ in the low-energy sector. We note that by taking only part of $\bar{H}_{R}$ and changing $s_x$ to $s_z$, we can model $H_{Rn}$. The Zeeman term $H_{z}$ is given by
\begin{equation}
\bar{H}_Z = \Delta_Z \sum_{i, \mu, \mu'}  c_{i\mu}^\dagger { s}_{x, \mu \mu'} c_{i\mu'}.
\end{equation}

{\it Spectrum with SOI.} A part of the Rashba SOI, $H_{Rx}$ ($H_{Rn}$), can be easily included in  $H_0$. The spin $s_x$ ($s_z$) is a good quantum number for the Hamiltonian $H_0+H_{Rx}$ ($H_0+H_{Rn}$).
In this case, the SOI only results in the spin-dependent shift of the momentum, $k_y \to k_y -s_x \alpha_R/\hbar \upsilon_F$ ($k_y \to k_y -s_z \alpha_{Rn}/\hbar \upsilon_F$),  see Fig. \ref{fig:spectrum_0}b.  
Next, we treat the remaining SOI terms, $H'=H_{so}+H_{Ry}$ ($H_{so} $), as a perturbation. We note that $H'$
($H_{so} $) is proportional to $\tau_z$. At the same time, the wavefunctions $\psi_m(x)$ [see Eq. (\ref{wavefunction_0})] are eigenstates of the Pauli matrix $\tau_1$, so the intrasubband matrix elements vanish, \cite{footnote_1} which is consistent with  Kramers degeneracy at $k_y=0$ for a time-reversal invariant Hamiltonian. 
The intersubband matrix elements $t_{mm'}$, however, are non-zero, but
they contain a strong suppression factor arising from the sublattice degree of freedom as follows. At small momenta, the mass term $\Delta \sigma_3$ dominates in the Hamiltonian, so the wavefunctions $\psi_m(x)$ are close to the eigenstate of  $\sigma_3$. As a result, the sublattice terms in $H'$ ($H_{so} $), $1-\sigma_3$ and $\sigma_1$, lead to a suppression of SOI effects, where the intrinsic SOI is suppressed by a factor $(E-\Delta)/\Delta \ll 1$ and the Rashba SOI by a factor $\sqrt{(E-\Delta)/\Delta}$.
Thus, the corrections to the spectrum are small in the parameter $t_{mm'}/\omega_{mm'}$, where $\omega_{mm'}$ denotes the subband splitting at given $k_y$. However, these terms lead to an {\it anticrossing} between two different subbands with opposite spin (with the same spin) along $x$ (along $z$), see Fig.~\ref{fig:spectrum_0}c. We note here that in spite of having strong SOI, the spin degeneracy is not lifted in case of quantum wires.

\begin{figure}[!tbp]
    \centering
    \includegraphics[width=0.75\columnwidth]{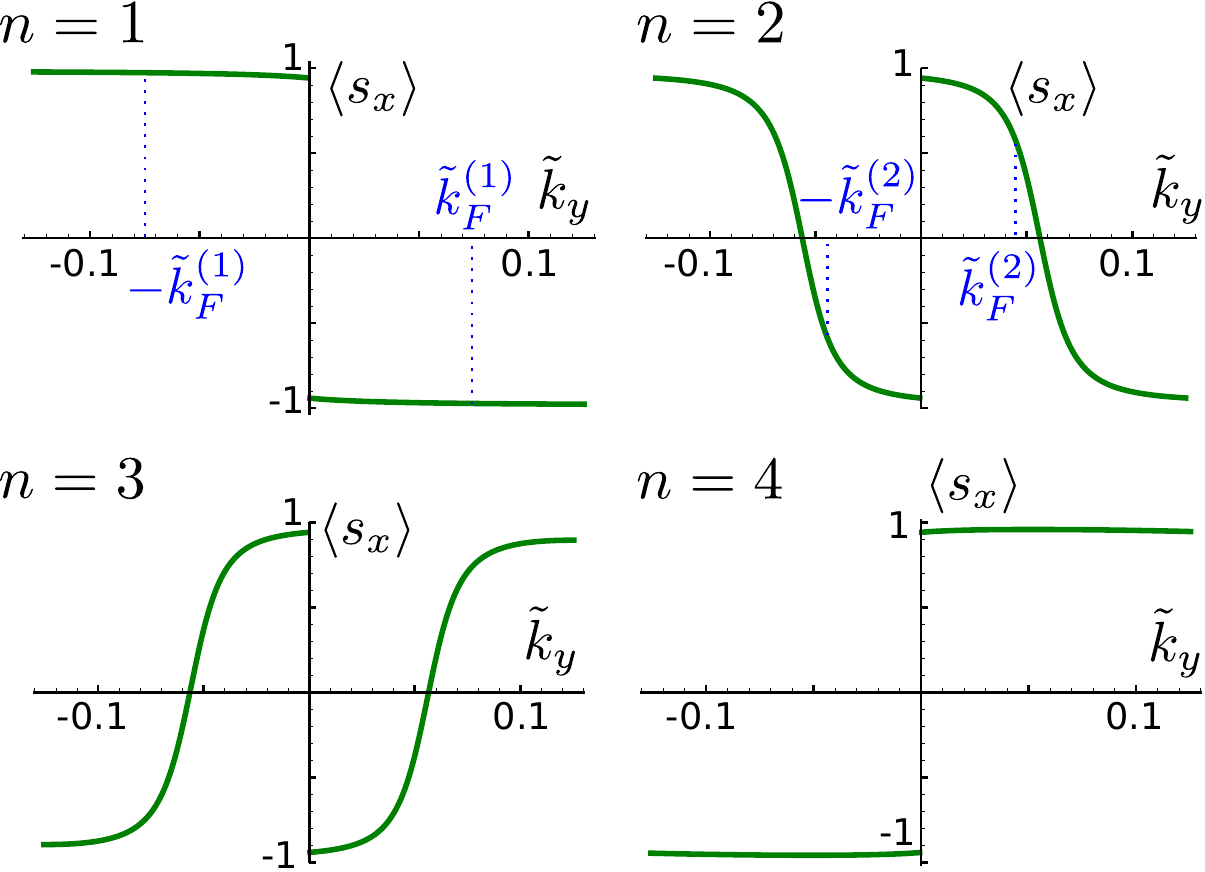}
\caption{The spin polarization $\left< s_x\right>$ along $x$ direction as function of  momentum $\tilde k_y$ for the $n$th level defined in Fig. \ref{fig:spectrum_0}c. The spin projections onto the $y$ and $z$ directions vanish. Away from the anticrossings the spin is almost perfectly aligned along $x$. If the chemical potential $\mu$, defining the Fermi momentum $\tilde k_{F}^{(n)}$ for the $n$th subband, is tuned close to the anticrossing (see Fig. \ref{fig:spectrum_0}c),  the total polarization $\left< s_x\right>$ of a left (right) propagating electron is non-zero.
}
\label{fig:Rashba_Full}
\end{figure}

{\it Helical modes via electric field.} The Rashba SOI induced by an electric field offers the possibility to generate helical modes in a time-reversal invariant system. As shown above, $H'$ results in subband anticrossings,
see Fig. \ref{fig:spectrum_0}c. Sufficiently far away from them, $t_{mm'}\ll \omega_{mm'}$, the subbands are spin-polarized by the Rashba SOI $H_{Rx}$ in the $x$ direction, see Fig. \ref{fig:Rashba_Full}. However, passing through the anticrossing  the spin polarization goes through zero and changes sign. All this suggests that if the Fermi level is tuned close to the anticrossing (see Fig. \ref{fig:spectrum_0}c) in such a way that there are four propagating modes (two left and two right), the system is in a quasi-helical regime. The lowest subband $n=1$ is almost fully spin-polarized and transports opposite spins into opposite directions, whereas the next subband $n=2$ is only partially polarized.
This means that  scattering due to impurities between subbands is allowed and helical modes are not protected from backscattering.

\begin{figure}[!bp]
    \centering
    \includegraphics[width=\columnwidth]{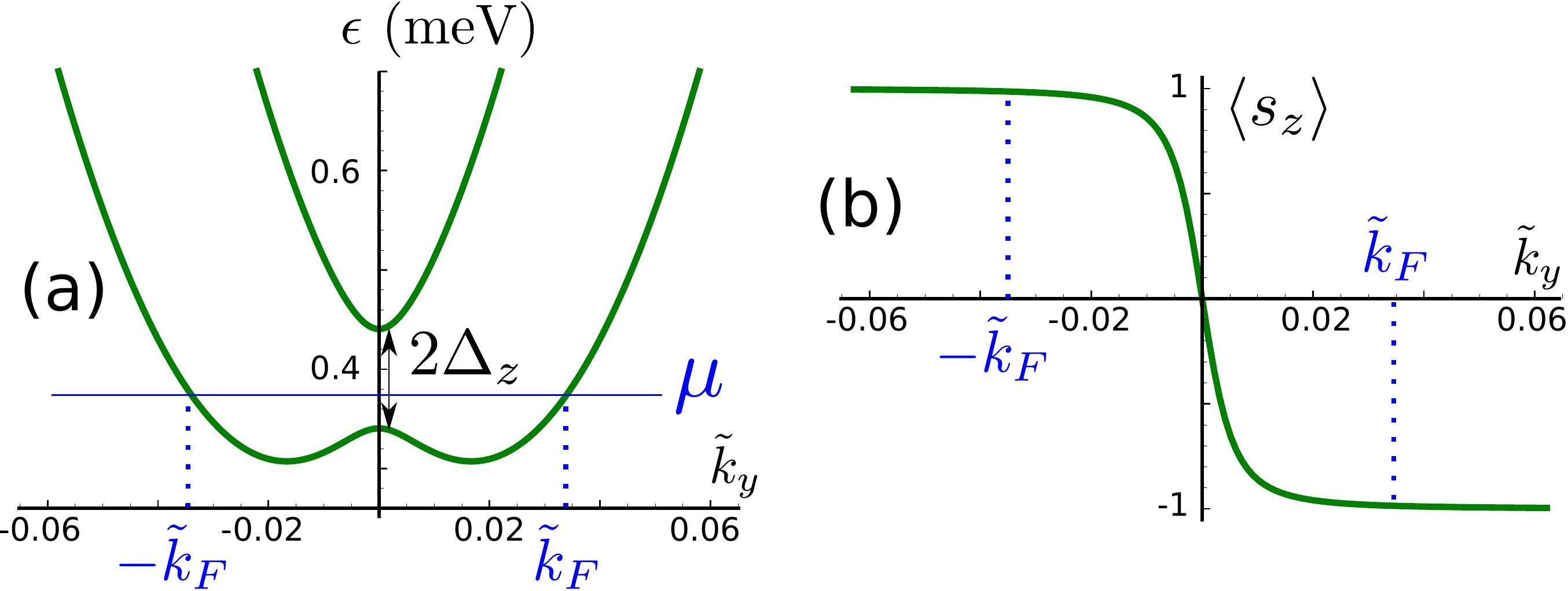}
\caption{ (a) The two lowest energy levels of $H_0+H_{Rn}+H_{so}+H_Z$, cf. Fig. \ref{fig:spectrum_0}. The Zeeman term lifts the Kramers degeneracy at $\tilde k_y=0$ and opens a gap $2\Delta_Z$. If $\mu$ is tuned inside the gap, the system is in a helical regime with a left (right) propagating mode with spin down (up).
(b) The spin polarization $\left< s_z\right>$ as function of the momentum $\tilde k_y$ for the lowest level.  The parameters are chosen as $W=50a$, $\alpha_{Rn}=10 \ {\rm meV}$, and $\Delta_Z =0.05 \ {\rm meV}$.
}
\label{fig:Rashba_Zeeman}
\end{figure}

{\it Helical modes via magnetic field.} If a Rashba SOI (along $x$) is generated by a spatially varying magnetic field, the time-reversal invariance of the system is broken, giving rise to a Zeeman term $H_Z$. The corresponding magnetic field, pointing along $z$, is perpendicular to the spin quantization axis determined by the Rashba SOI. 
Thus, the spin degeneracy at $k_y=0$ gets lifted, and a gap of size $2\Delta_Z$ is opened, see Fig. \ref{fig:Rashba_Zeeman}. The spin polarization along $z$
is given by
\begin{equation}
\left< s_z \right> =\frac{\omega_{\downarrow \uparrow }}{\sqrt{\omega_{\downarrow \uparrow } + 4\Delta_Z^2}},
\label{eq:pol}
\end{equation}
where $\omega_{\uparrow \downarrow}$ is the energy difference between spin up and spin down states at given momentum $k_y$ for the unperturbed problem $H_0+H_{Rn}+H_{so}$. \cite{footnote2} If the chemical potential $\mu$ is tuned inside the gap,
 there is one  mode propagating to the left and one to the right. Moreover, these two modes carry opposite spins with almost perfect polarization, $\left|\left< s_z \right>\right|\approx 1$, provided that $\Delta_Z \ll 16 \alpha_{Rn}^2/\Delta$. Thus, the system is in a helical regime.

{\it Majorana fermions.} Helical modes as in Fig. \ref{fig:Rashba_Zeeman} have attracted considerable attention in various candidate systems not only as a platform for spin-filters \cite{streda}  but also as a platform for generating MFs.~\cite{alicea_review_2012}
MFs are particles that are  their own antiparticles. When the quantum wire is brought into tunnel contact with an $s$-wave superconductor inducing a proximity gap
$\Delta_{sc}$, states with opposite spins and momenta are coupled giving rise to an effective $p$-wave pairing.
In the topological phase, MFs emerge as midgap boundstates, one localized at each end of the quantum wire. This phase emerges if 
$\Delta_Z^2 > \Delta_{sc}^2 + { \mu}^2$ is satisfied, where 
$\mu$ is now counted from the middle of the gap. Since the derivation is similar to previously studied cases,~\cite{MF_wavefunction_klinovaja_2012,MF_CNT_2012, nanoribbon_KL,Two_field_Klinovaja} we defer the details to App. ~\ref{Supp}.
The $\rm MoS_2$ monolayer quantum wires offer the unique possibility to probe MFs not only by transport but also by optical spectroscopy.

{\it Quantum dots.} In contrast to gapless graphene, \cite{Guido_Nature,Guido_dots} quantum dots \cite{Spin_qubits, kloeffel_prospects_2013} in $\rm MoS_2$ can be created by gates. \cite{Lieven_exp, Amir_exp} We note that the confining potential should be sharp enough to lift the valley degeneracy, as shown above for the quantum wires, and, in addition, to insure a non-equidistant spectrum, which is more suitable for optical experiments. If vanishing boundary conditions are imposed also along $y$, the momentum $k_y$ is quantized, $\kappa_{yn}=\pi n/(W_y+2a/\sqrt{3})$, where $W_y$ is width in the $y$ direction, and $n$ is a positive integer. The dot spectrum then becomes
$E_{n,m}=\sqrt{(\hbar \upsilon_F)^2 (\kappa_m^2+ \kappa_{yn}^2) +\Delta^2}$.
Each level is spin degenerate and the intrinsic SOI can neither lift this Kramers degeneracy, nor,  due to its symmetry, change substantially splittings between levels [in the small parameter $\alpha (E-\Delta)/\Delta\ll \alpha$, see above]. The spin degeneracy can be lifted with a magnetic field, say, along $z$. Similar to nanotubes,~\cite{klinovaja_cnt} EDSR can then be achieved by applying an oscillatory electric field $E$ also along $z$, causing $\alpha_R$ to oscillate and thereby inducing spin rotations at a Rabi frequency $\sim |\alpha_R|$.
Thus, we conclude that quantum dots in $\rm MoS_2$ host well-defined Kramers doublets that
can serve as platform for spin qubits.~\cite{Spin_qubits, kloeffel_prospects_2013}

We acknowledge stimulating discussions with Parisa Fallahi, Richard Warburton, and Dominik Zumbuhl. This work is supported by the Swiss NSF, NCCR Nanoscience, and NCCR QSIT.

\appendix

\section{Majorana Fermions \label{Supp}}

We give here more details of the derivation of the Majorana fermions (MFs) introduced in the main text. Thereby we  closely follow the derivation given in Ref.~\onlinecite{MF_wavefunction_klinovaja_2012} 
which requires a few minor modifications for the present case.
If the chemical potential $\mu$ is tuned inside the gap $2\Delta_Z$ opened by the magnetic field ${\bf B}_n$ at $k_y=0$,  the two propagating modes are helical, see Fig. 4. 
The same helical states can be obtained by a Rashba SOI induced by an electric field in the presence of a uniform magnetic field giving rise to a Zeeman splitting $2\Delta_Z$.
 If such a quantum wire is brought into tunnel contact with an $s$-wave superconductor, a superconducting proximity gap $\Delta_{sc}$ is induced in the wire. Through the  pairing mechanism coupling Kramers partners,
 the helical states get paired into a $p$-wave-like superconducting state.~\cite{Sato,lutchyn_majorana_wire_2010,oreg_majorana_wire_2010,alicea_review_2012}
There are no propagating modes inside the gap but there could exist boundstates localized at the ends of the wire. If a certain topological criterion is satisfied, these states are MFs, particles that are their own antiparticles. To find this criterion, we describe the system by an effective linearized model for the exterior ($\chi=e$, states with momenta close to the Fermi momentum, $k_e=k_F$) and the interior branches ($\chi=i$, states with  momenta close to $k_i=0$). \cite{MF_wavefunction_klinovaja_2012} The electron operator is 
represented as
$\Psi(y)=\sum_{\rho={\pm 1},\chi=e,i} e^{i \rho k_\chi y} \Psi_{\rho\chi }$,
where the sum runs over the right ($R$, $\rho=1$) and left ($L$, $\rho~=~-1$) movers and $\Psi_{\rho\chi }$ is an annihilation operator for the $(\rho,\chi )$ branch of the spectrum.
The effective Hamiltonian becomes
\begin{align} 
&H = -i \hbar \upsilon  \rho_3 \chi_3 \partial_y  + \Delta_Z \eta_3 \rho_1 (1+\chi_3)/2 \nonumber\\
&\hspace{20pt}+ { \Delta}_{sc}  \eta_2 \rho_2 (1+\chi_3)/2 + {\bar \Delta}_{sc}  \eta_2 \rho_2 (1-\chi_3)/2.
\label{static}
\end{align}
in the basis
$$\widetilde{\Psi}=(\Psi_{Re},\Psi_{Le},\Psi_{Re}^\dagger,\Psi_{Le}^\dagger,\Psi_{Li},\Psi_{Ri},\Psi^\dagger_{Li},\Psi^\dagger_{Ri}),$$
where the Pauli matrices $\chi_i$ ($\eta_i$) act in the interior-exterior branch (electron-hole) space.

Here,  $\Delta_Z = g\mu_B B_n/2$ is the Zeeman energy, and $\upsilon = (\partial E/\partial \hbar k_y)|_{k_y=k_F}$ is the velocity at the Fermi level.
In the limit of strong Rashba SOI ($\alpha_{Rn} \gg \Delta_Z, \Delta_{sc}$), the strength of the effective proximity induced superconductivity acting on the exterior branches $\bar \Delta_{sc}$ due to the nearly perfect spin polarization at the Fermi wavevector $k_F$ is equal to $\Delta_{sc}$. In the opposite limit of  weak Rashba SOI ($\alpha_{Rn} \ll \Delta_Z$), $\bar \Delta_{sc}$ is getting suppressed by the magnetic field, $\bar \Delta_{sc} = \Delta_{sc} k_{n}/k_F $, where $k_n = 2\pi/\lambda_n$ ($k_n = 2\alpha_R/\hbar \upsilon_F$) for Rashba SOI induced by rotating magnetic  fields (by electric fields).
Note that the Fermi wavevector $k_F$ grows with magnetic field as $k_F \propto \sqrt{\Delta_Z}$.

All this together leads us to  the  criterion for the topological phase  given by 
\begin{align}
\Delta_Z>\sqrt{\Delta_{sc}^2+\mu^2 },
\end{align}
where the chemical potential $\mu$ is now calculated from the middle of gap $2\Delta_Z$.

Similarly, following Refs. \onlinecite{MF_wavefunction_klinovaja_2012, Two_field_Klinovaja}, we can also obtain the  localization length of the MFs. For example, in the strong SOI regime and for $\mu=0$, the wavefunction of the left localized MF is written in the basis ${\bar \Psi} = ({\Psi}_\uparrow, {\Psi}_\downarrow, {\Psi}_\uparrow^\dagger, {\Psi}_\downarrow^\dagger)$ as
\begin{align}
\varPhi_{M}(y)=\begin{pmatrix}
i\\
1\\
-i\\
1
\end{pmatrix}   e^{-k_-^{(i)}y} -
\begin{pmatrix}
i\ e^{ik_F y}\\
 e^{-ik_F y}\\
-i\ e^{-ik_F y}\\
 e^{ik_F y}
\end{pmatrix}e^{-k^{(e)} y},\label{wave_MF}
\end{align}
where $k_-^{(i)} = (\Delta_Z - \Delta_{sc})/ \hbar \upsilon $ and $k^{(e)} =  \Delta_{sc}/ \hbar \upsilon$ \cite{MF_wavefunction_klinovaja_2012}. The localization length is determined by the smallest gap in the system $\xi = {\rm max} \{1/k_-^{(i)}, 1/k^{(e)}\}$. Here, ${\Psi}_{\uparrow(\downarrow)}^\dagger$ is a creation operator of the electron with spin up (down), where the spin quantization axis is determined by the Rashba SOI.
In the weak SOI regime deeply in the topological phase, the left localized MF wavefunction is written as
\begin{equation}\varPhi_{M}(y)=
\begin{pmatrix}
e^{-i\pi/4}\\
i e^{i\pi/4}\\
e^{i\pi/4}\\
-i e^{-i\pi/4}
\end{pmatrix} \sin (k_F y) e^{-\bar{k}^{(e)} y},
\label{MF_weak_SOI}
\end{equation}
where $\bar k^{(e)} = \bar \Delta_{sc}/ \hbar \upsilon$ \cite{MF_wavefunction_klinovaja_2012}. Note that both solutions explicitly satisfy the MF condition of being self-conjugate, $\varPhi_{M}\cdot{\bar \Psi}
=[\varPhi_{M}\cdot{\bar \Psi}]^\dagger$.


\begin{thebibliography}{100}
\bibitem{footnote_1} More generally, the intersubband matrix element $t_{mm'}$ vanishes if $m-m'=2l$, where $l$ is an integer.
\bibitem{footnote2} We note that the spin polarization close to the anticrossing is also described  by Eq. (\ref{eq:pol}), where we make a change $\Delta_Z \to t_{ij}$, see Fig. \ref{fig:Rashba_Full}.
\end{thebibliography}
\end{document}